\documentclass{sig-alternate}

\usepackage[latin1]{inputenc} % entree 8 bits iso-latin1
\usepackage[T1]{fontenc}      % encodage 8 bits des fontes utilisees
\usepackage{url}

\begin{document}

\title{On Benchmarking Embedded Linux Flash File Systems}
%% \subtitle{[Extended Abstract]
%% \titlenote{A full version of this paper is available as
%% \textit{Author's Guide to Preparing ACM SIG Proceedings Using
%% \LaTeX$2_\epsilon$\ and BibTeX} at
%% \texttt{www.acm.org/eaddress.htm}}}

\numberofauthors{3} %  in this sample file, there are a *total*
\author{
% 1st. author
\alignauthor
Pierre Olivier\\
       \affaddr{Universit\'e Europ\`eenne de Bretagne, France}\\
       \affaddr{Universit\'e de Brest,}\\
       \affaddr{CNRS, UMR 3192 Lab-STICC,}\\
       \affaddr{20 avenue Le Gorgeu, 29285 Brest cedex 3, France}\\
       \email{pierre.olivier@univ-brest.fr}
% 2nd. author
\alignauthor
Jalil Boukhobza\\
       \affaddr{Universit\'e Europ\`eenne de Bretagne, France}\\
       \affaddr{Universit\'e de Brest,}\\
       \affaddr{CNRS, UMR 3192 Lab-STICC,}\\
       \affaddr{20 avenue Le Gorgeu, 29285 Brest cedex 3, France}\\
       \email{boukhobza@univ-brest.fr}
% 3rd. author
\alignauthor
Eric Senn\\
     \affaddr{Universit\'e Europ\`eenne de Bretagne, France}\\
       \affaddr{Universit\'e de Bretagne Sud,}\\
       \affaddr{CNRS, UMR 3192 Lab-STICC,}\\
       \affaddr{C.R. C Huygens, 56321 Lorient, France }\\
       \email{eric.senn@univ-ubs.fr}
%~ % 4th. author
%~ \alignauthor Lawrence P. Leipuner\\
       %~ \affaddr{Brookhaven Laboratories}\\
       %~ \affaddr{Brookhaven National Lab}\\
       %~ \affaddr{P.O. Box 5000}\\
       %~ \email{lleipuner@researchlabs.org}
%~ % 5th. author
%~ \alignauthor Sean Fogarty\\
       %~ \affaddr{NASA Ames Research Center}\\
       %~ \affaddr{Moffett Field}\\
       %~ \affaddr{California 94035}\\
       %~ \email{fogartys@amesres.org}
% 6th. author
%~ \alignauthor Charles Palmer\\
       %~ \affaddr{Palmer Research Laboratories}\\
       %~ \affaddr{8600 Datapoint Drive}\\
       %~ \affaddr{San Antonio, Texas 78229}\\
       %~ \email{cpalmer@prl.com}
}
% There's nothing stopping you putting the seventh, eighth, etc.
% author on the opening page (as the 'third row') but we ask,
% for aesthetic reasons that you place these 'additional authors'
% in the \additional authors block, viz.
%~ \additionalauthors{Additional authors: John Smith (The Th{\o}rv{\"a}ld Group,
%~ email: {\texttt{jsmith@affiliation.org}}) and Julius P.~Kumquat
%~ (The Kumquat Consortium, email: {\texttt{jpkumquat@consortium.net}}).}
\date{}
% Just remember to make sure that the TOTAL number of authors
% is the number that will appear on the first page PLUS the
% number that will appear in the \additionalauthors section.

\maketitle

\begin{abstract}
Due to its attractive characteristics in terms of performance, weight and power consumption, NAND flash memory became the main non volatile memory (NVM)  in embedded systems. Those NVMs also present some specific characteristics/constraints: good but asymmetric I/O performance, limited lifetime, write/erase granularity asymmetry, etc.

Those peculiarities are either managed in hardware for flash disks (SSDs, SD cards, USB sticks, etc.) or in software for raw embedded flash chips. When managed in software, flash algorithms and structures are implemented in a specific flash file system (FFS). In this paper, we present a performance study of the most widely used FFSs in embedded Linux: JFFS2, UBIFS,and YAFFS. We show some very particular behaviors and  large performance disparities for tested FFS operations such as mounting, copying, and searching file trees, compression, etc.

\end{abstract}

%% Categories
{Measurements}

% Mots-clef
 % NOT required for Proceedings

\section{Introduction}

NAND and NOR flash are the most common types of flash memories \cite{woodhouse_jffs:_2001}. NOR Flash memory provides good read performance and random byte access at the cost of slow write operations and low data densities. It is suitable for code storage and execution and is used as a replacement of DRAM is some mobile appliances. NAND flash memory is dedicated to data storage, it provides more balanced read and write performance (even though asymmetric) and a higher data density, at a lower cost. It is used as the main secondary storage for many embedded systems. In this paper we are only concerned with NAND flash memories (designated as flash memory in the rest of the paper).

Data in flash memory is organized hierarchically : a chip is divided into \textit{planes}, themselves divided into \textit{blocks}. Blocks are composed of \textit{pages}. Finally, pages can be divided into \textit{user-data space}, and a small meta-data section called the \textit{out-of-band area}. Today, flash blocks typically contain blocks with 64 pages. These page size is generally 2048 bytes \cite{woodhouse_jffs:_2001}.

Flash memory supports 3 operations : \textit{read} and \textit{write} operations performed at a page level, and the \textit{erase} operation on an entire block. As flash memory provides many benefits, it also comes with specific drawbacks due to its internal intricacies. First, the erase-before-write limitation which imposes a costly block erase operation before writing a data. The consequence of this constraint is the inability to achieve efficient in-place data modification.The other very important drawback is the limited lifetime. A flash memory cell can sustain a limited number of erase operations, above which it can no more retain data. Typically, a NAND flash  cell has an endurance estimated between $10^4$ and $10^5$ write/erase cycles \cite{gal_algorithms_2005}. Moreover, NAND flash cells can leave factory with faulty cells. Flash management algorithm partly resolve the problem by providing spare cells and implementing prevention mechanisms.

Flash memory constraints require a specific management. The erase-before-write rule is bypassed by writing new versions of data to other locations and  invalidating old ones. Such a mechanism must involve a garbage collector that periodically recycles invalidated data into free space. Moreover, it is necessary to evenly distribute the write/erase cycles (wear out) over the whole flash memory in order to maximize its lifetime. This is called \textit{Wear leveling}.

Specific flash memory algorithms and structures can be managed through a hardware layer called the Flash Translation Layer (FTL) \cite{gal_algorithms_2005, gupta_dftl_2009}. The FTL layer implements the wear leveling and garbage collection algorithms and is used in Solid State Drives (SSD), compact flash, USB sticks, etc. Performance of FTLs is a very challenging topic and several studies have been done in the domain. Flash memory algorithms and structure are generally implemented in software when the flash is integrated into an embedded system. In embedded Linux operating system, this is performed through a dedicated FFS \cite{egger_file_????, woodhouse_jffs:_2001, adrian_hunter_brief_2008, manning_how_2010}. 

Embedded Linux market explosion lead us to more seriously consider FFS performance issues. In our opinion, too few studies were done on the performance of those very widely used FFS. This paper is an attempt to partially fill this gap by presenting a testing methodology and results on most used FFSs. This work focuses on specific file system operations such as mounting, copying file trees, file searches, and file system compression. We do not consider flash specific operations such as sequential and random read/write operations and some specific garbage collection and wear leveling tests which will be considered in a future study.

In the next section we present some related work about FFS performance evaluation. Then, we roughly explain FFSs features, and provide some examples. Next we describe the benchmarking methodology, and then we discuss the results before concluding.

\section{Related work}

In \cite{liu_analysis_2010}, the authors compared JFFS2, YAFFS2 and UBIFS. Analyzed metrics are file system mount times, system performances (using the \textit{Postmark} benchmark), memory consumption, wear leveling and garbage collection cost. They conclude that the choice of an FFS can be motivated by hardware constraints, particularly available flash size. In \cite{michaelopdenacker_update_2008} the authors compared the performance of JFFS2, YAFFS, UBIFS and SquashFS, a read-only file system which is not dedicated to flash memory. It is stated that UBIFS is the way to go in case of large flash chips. For lower sizes, JFFS2 is favored as compared to YAFFS2.
  
A performance evaluation of UBI and UBIFS is provided in \cite{toshiba_corporation_evaluation_2009}. The authors identified very specific weaknesses of the file system about mount time and flash space overhead. In \cite{elinux_ffs_2012}, benchmarks are performed on JFFS2, YAFFS2 and UBIFS, comparing mount time, memory consumption and I/O performance on various kernel versions from 2.6.36 to 3.1.

Our study differs from the above related work  in that it gives more details on the performance behaviors of FFS on the mounting and compression performance, in addition to new operation benchmarking (file tree copying, file search, and compression). In our point of view, FFS performance evaluation can be split into two parts: 1) the performance of the FFS and how it accesses to its meta data (that can be stored in the flash itself), and 2) the performance of accessing the flash memory throughout the chosen FFS by applying different I/O workloads. In this paper, we focus only on the first part.

\section{Linux Flash File Systems}
In this section we briefly present three recent NAND FFSs. Before going into further details about Linux FFSs, we introduce the encompassing software architecture.

   \subsection{A layered organization in the kernel}
   
   \begin{figure}
     \centering
     \includegraphics[width=0.33\textwidth]{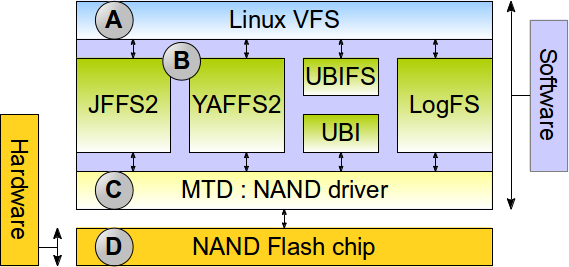}
     \caption{FFS into the linux kernel layers.}
     \label{Layers}
   \end{figure}
   
   The location of FFSs into the kernel is depicted on Figure \ref{Layers}. We can identify several software layers : VFS, the various FFSs layers, and the Memory Technology Device (MTD).
   
   The high-level \textit{Virtual File System} layer (A on Figure \ref{Layers}) is located above the FFS one. VFS allows different file systems to coexist, and present the same interface to the user level. This implies that each file system written for Linux must be compliant with the VFS interface. VFS uses memory structures created on demand : when needed, VFS asks the file system (B on Figure \ref{Layers}) to build the corresponding object. VFS also maintains cache systems to accelerate I/O operations. At a lower level, the FFS must be able to perform raw flash I/O operations. The NAND driver is provided by the \textit{MTD} \cite{_linux_????} layer (C on Figure \ref{Layers}), a generic subsystem which role is to provide a uniform interface to various memory devices, comprising NAND and NOR flash memory.

    \subsection{FFSs algorithms and structures}
    Linux FFSs studied in this paper are depicted in Figure \ref{Layers}.

    There are some common features to those FFS: 1) for instance, the data \textit{compression} is supported by most of them. Compression not only reduces the size of the data on the media, but it also reduces I/O load when performed on the fly (i.e. at runtime) at the expense of an increase of CPU load (for compression/decompression). 2) \textit{Bad block management} service is also provided by all FFSs. Bad blocks are generally identified using a specific marker in the out-of-band area, and never used. 3) All the FFSs provide wear leveling and garbage collection mechanisms. 4) Finally, some FFS also provide journaling capabilities, which consists in writing in a journal the description of each file system modifications before performing the modification itself. The purpose of this service is to keep a valid data version to use in case of a system crash. The modification is therefore performed out-of-place, and validated on the journal once completely performed. UBIFS, for instance, is a journaled FFS. 
    
    %UBIFS and LogFS are journalized. 

%%%%%%%%%    
%    The impossibility to perform in-place data updates in flash memory makes it a a good candidate for \textit{log-structured} and \textit{journaling} file systems \cite{guido_r._kok_flash_2008}. Journaling file systems write a small amount of meta-data for each file system modifications into a journal. After a system crash (for example a power failure) the journal is scanned, and the last journalized modifications are validated or invalidated according to some specific rules. In a \textit{log-structured} file system, the file system meta-data \textit{and} data are appended to the log in a sequential way. This was invented to avoid rotational delays in traditional hard drives, and is useful for other reasons in flash memory. %%% quelles raisons %%%
%%%%%%%%%
    
   %  The next section briefly depicts some recent FFS features.
    
    \subsubsection{JFFS2}
    
    The \textit{Journaling Flash File System version 2} (JFFS2) \cite{woodhouse_jffs:_2001} is today's most commonly used FFS. It has been present in the kernel mainline since Linux 2.4.10 (2001). The storage unit for data/meta-data is the JFFS2 \textit{node}. A node represents a file, or a part of a file, its size varies between a minimum of one flash page and a maximum of half an erase flash block. At mount time, JFFS2 has to scan the entire partition in order to create a direct mapping table to JFFS2 node on flash.
    
    JFFS2 works with three lists of flash memory blocks : 1) the \textit{free} list is a list of blocks that are ready to be written. Each time JFFS2 needs a new block to overwrite a node, the block is taken from this list. 2) the \textit{clean} list contains blocks with valid nodes, and 3) the \textit{dirty} list contains blocks with at least one invalid node. When the free space becomes low, the garbage collector erase blocks from the dirty list. In order to perform wear leveling, JFFS2 occasionally (with a given probability) chooses to pick one block from the clean list instead (copying its data elsewhere beforehand).
    
    Although JFFS2 is widely used, it scales linearly according to the flash size from a performance point of view (more details on the following sections). This means JFFS2 RAM usage and mount time increase linearly with the size of the managed flash device. JFFS2 supports multiple compression algorithms comprising \textit{Zlib} and \textit{Lzo}. Nodes are (de)compressed at runtime when they are accessed.
    
    \subsubsection{YAFFS2}
    
    The original specifications of \textit{Yet Another Flash File System} (YAFFS version 2) \cite{manning_how_2010} dates back to 2001. The integration of YAFFS2 into the kernel is done with a patch.

     %%%%%% Journalisé ou pas %%%%%%
    
    YAFFS2 stores data in a structure called the \textit{chunk}. Each file is represented by one \textit{header} chunk containing meta-data (type, access rights, etc.) and data chunks containing file/user data. The size of a chunk is equal to the size of the underlying flash page. Chunks are written on flash memory in a sequential way. In addition to meta-data stored in the header chunk, YAFFS also uses the out-of-band area of data chunks, for example, to invalidated updated data. Like JFFS2, the whole YAFFS partition is scanned at mount time. YAFFS does not support compression. 
    
      Garbage collection (GC) can be performed either 1) when a write occurs and the block containing the old version is identified as completely invalid, it is then erased by GC, or 2) when the free space goes under a given threshold, GC selects some blocks containing valid data, copies still valid chunks to another locations and erases the block.
    
    Like JFFS2, YAFFS2 performance and meta data size scales linearly according to flash memory size. 
 
    \subsubsection{UBI + UBIFS}
    \textit{UBIFS} \cite{adrian_hunter_brief_2008} is integrated into the kernel mainline since Linux 2.6.27 (2008). UBIFS aims to solve many problems related to the scalability of JFFS2. UBIFS relies on an additional layer, the UBI layer. UBI \cite{gleixner_ubi-unsorted_2006} performs a logical to physical flash block mapping, and thus discharges UBIFS from the wear leveling and bad block management functions. 

    %%%%%% Journalisé ou pas %%%%%%

While JFFS2 and YAFFS2 use tables, UBIFS uses tree-based structures for file indexing. The index tree is stored on flash through \textit{index nodes}. The tree leaves point to flash locations containing flash data or meta-data. UBIFS also uses standard \textit{data nodes} to store file data. UBIFS partitions the flash volume into several parts: the \textit{main area} containing data and index nodes, and the \textit{Logical erase block Property Tree} area containing meta-data about blocks: erase and invalid counters used by the GC. As flash doesn't allow in-place data updates, when updating a tree node, the entire parent and ancestors nodes are moved to another location, that is why it is called a \textit{wandering tree}.

%As flash doesn't allow in-place data updates, when updating a tree node, the entire node must be rewritten to another location. The parent node must also be updated to point to the new child. As for its parent, and so on up to the root node. Such a management algorithm make the tree a \textit{wandering tree}.
    
    In order to reduce the number of  flash accesses, file data and meta-data modifications are buffered into the main memory and periodically flushed on flash. Each modification of the file system is logged by UBIFS in order to maintain the file system consistency in case of a power failure. UBIFS supports the \textit{Lzo} and \textit{Zlib} compression algorithm, and provides a \textit{favor Lzo} compression option, using alternatively Lzo and Zlib for a more balanced CPU usage ratio.
    
    With the use of tree structures, UBIFS layer performance scales in a logarithmic way with the size of the underlying flash partition, while the UBI layer scales linearly \cite{_linux_????}. 
    
  %  \subsubsection{LogFS}
   % \textit{LogFS} \cite{engel_logfs-finally_2005, guido_r._kok_flash_2008} is present in the kernel mainline since Linux 2.6.34 (2010). Like UBIFS, LogFS uses a tree-based structure for file indexing. LogFS divides the flash memory into sections, notably the \textit{journal}, containing the on-flash position of the root node of the index tree, and the \textit{object store} containing all file system file data and meta data. At mount time, the journal is scanned to find out the last version of the root node. The index tree can then be rebuilt. As each file update generates an update of the root node, the journal is buffered in RAM. The journal can be moved to different flash locations to avoid rapid wearing of the block containing it. Garbage collection in Logfs is a complex process which causes slowdowns when the occupation ratio of the partition becomes high \cite{engel_garbage_2007, guido_r._kok_flash_2008}.

%%% scales comment ?%%%%

\section{Benchmarking methodology}

We performed our benchmarking on an \textit{Armadeus} APF27 embedded board, equipped with a i.MX27 CPU clocked at 400 Mhz, 2*64 MB of RAM, and a 256 MB SLC (Single Level Cell) NAND flash memory chip used for secondary storage. The flash chip is a Micron SLC NAND flash with a reported read latency of 25 $\mu$s, a write latency of 300 $\mu$s, and an erase latency of 2 ms. The used Linux kernel version is 2.6.38.8. % LogFS was already integrated in the mainline in this version, but due to a large instability (crashing the kernel) caused by its usage, we do not provide results related to it.

\subsection{Performance metrics}

When performing complete FFSs benchmarking, one has to consider traditional file systems metrics : read and write I/O performance ; RAM usage, and its evolution according to the partition size, CPU usage, mount time, tolerance to power failures, compression etc. We have also to consider FFSs specific metrics, related to wear leveling, garbage collection, and bad block management. For space reasons, we only focus on high level file manipulations in this paper. We do not explicitly take into account FFS specific metrics. The considered metrics are: (un)mount time, operations (read, search and copy) on file trees, and compression impact on file system performances.

\paragraph{Execution time measurement and file tree generation}
\begin{figure}
  \centering
  \includegraphics[width=0.20\textwidth]{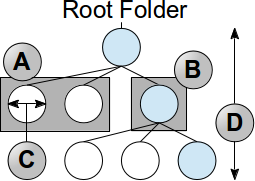}
  \caption{File tree generation parameters.}
  \label{arbo}
\end{figure}
We used the  the \verb+gettimeofday()+ system call to measure execution times. It provides a microsecond precision. The executed commands were launched with the help of the \verb+system()+ function.
In our benchmarks we used various file trees to efficiently measure FFSs performance. To do so, we developed a file tree generation tool that can be used on whatever file system. We lie upon several parameters to define the file tree as depicted in figure \ref{arbo}: the number of files per generated directory (A), the number of directories per generated directory (B), the size of generated files (C), and the file tree depth (D). The tool is very flexible as it allows to define those parameters with some probability distributions.

\paragraph{Benchmarking scenarios}\label{test}
We wrote some shell scripts performing various file system related operations. We measured the execution time of each operation with the method presented above. For the operations involving file tree manipulation, we define these file trees with the parameters presented earlier. Here we present two scenarios through which many operations were measured. 

In the \textit{scenario 1} (S1), we first prepared several FFSs images, varying compression options when available. The images are based on the same directory tree, which is a standard embedded Linux root file system. This tree contains 213 directories and 1122 files. Each directory of this rootfs contains a average of 4,95 files and 1 directory. Average file size is 13 KB. Each image is mounted in a 100 MB partition, and a recursive \verb+ls -R+ command is performed on the mount point to evaluate the FFS meta-data read operations. This forces VFS to ask the FFS for meta-data about file tree organization and file / directory names. A second \verb+ls -R+ is performed to isolate the cache effect of VFS. In the mounted partition we then create another file tree. The parameters for this file tree are a depth of 5, a number of files per generated directory obtained by a random normal distribution of mean value 4 and a standard deviation of 1 (\textit{norm(4, 1)}), a number of directories per generated directory of \textit{norm(3, 1)}, and a file size of \textit{norm(1024, 64)} bytes. The partition is then unmounted.

\begin{figure}
  \centering
  \includegraphics[width=0.30\textwidth]{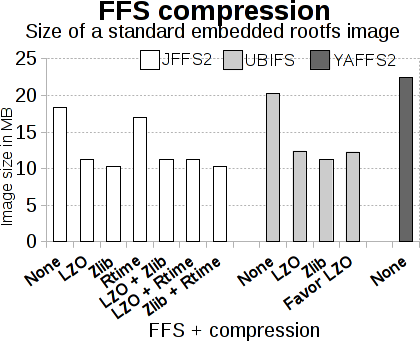}
  \caption{Compression efficiency}
  \label{compressionefficiency}
\end{figure}

In the \textit{scenario 2} (S2), we erase a 100 MB flash partition and create an empty file system. Next we create a file filling randomly the whole partition. This file is then destroyed. This forces the FFS to invalidate corresponding data, giving more representative conditions of the flash state for the rest of the scenario (this can be considered as a partition warm up). Then we create a file tree with the following parameters : a depth of 5, a variable number of files per generated directory, a file size of 750 bytes, and 2 directories per generated directory. We unmount the partition, then remount it. Next we perform a \verb+find+ command in the mount point, searching for a file that is not present in the file tree, in order to have the worst conditions (search the whole meta data). The whole file tree is then deleted, and the file system unmounted. We launched this scenario for each of the tested FFSs, with default compression options.

%%%%%%%%%%%%%%%%%%%%% A REVOIR
The first scenario allows to measure the compression efficiency on the file system, the mount time, meta data read (throughout the \verb+ls+ command), file tree creation on a given FFS, and the unmount time.
The second scenario allows to perform a warm-up of the flash partition (by making it dirty) before measuring the mount and unmount operations, the search on FFS meta-data (find command), and file tree deletion.
%%%%%%%%%%%%%%%%%%%%%

\section{Results and discussion}

\subsection{Compression}

  \paragraph{Compression efficiency}
  The size of S1 images is presented in Figure \ref{compressionefficiency}. We can observe that compression reduces the size of the stored data by up to 40\% in some cases for JFFS2 and UBIFS. Various FFS uncompressed images sizes are different because of the granularity of the file write operation that is different in each FFS. For example, JFFS2 manages nodes which size vary between 1 page and half of a flash block, while YAFFS2 always uses one page per chunk in addition to a dedicated page for the file header chunk (which is very space consuming mainly when we deal with small files). This explains the larger size of YAFFS image.
  
  \paragraph{Compression impact on S1}
  
  \begin{figure}
  \centering
  \includegraphics[width=0.45\textwidth]{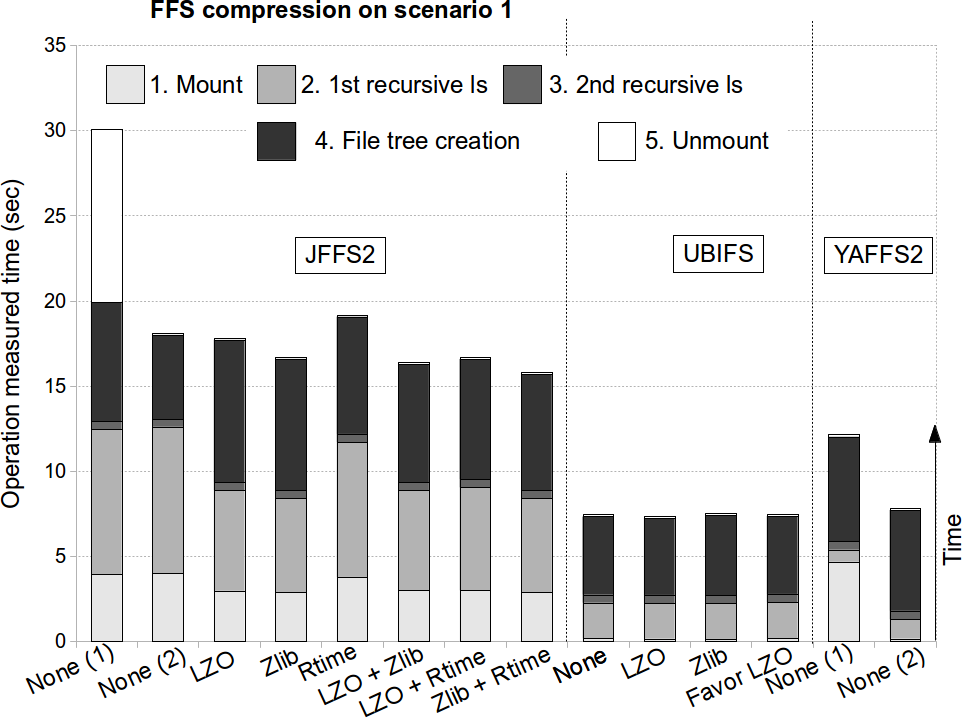}
  \caption{Scenario 1 operations execution times}
  \label{scenario1}
\end{figure}

\begin{figure*}
\centering
  \includegraphics[width=0.9\textwidth]{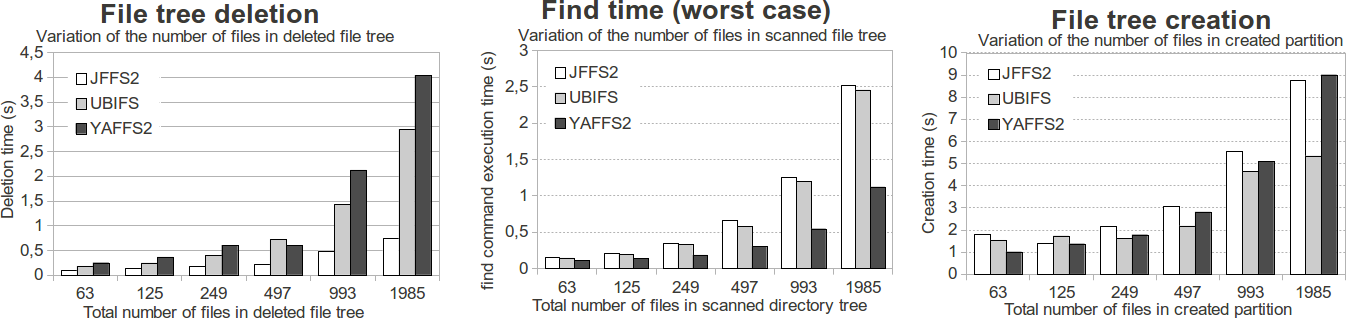}
  \caption{File system operation execution time results: find, mount, and file tree creation time in S2.}
  \label{filemanip1}
\end{figure*}

Figure \ref{scenario1} depicts the measured execution times of S1 operations. Before flashing a new image we perform a full flash erase, except between \textit{None(1)} and \textit{None(2)} for both YAFFS and JFFS2, which represent two consecutive runs of the S1 on the same uncompressed image. Compression reduces the stored data size and also flash I/Os, thus enhancing the overall performances (mount time, file tree creation, and ls command). In particular, for JFFS2's mount time and file tree creation time. We can also notice that UBIFS compression does not affect performance.

\subsubsection{File system operations}

\paragraph{Mount / Unmount times}
From Figure \ref{scenario1}, one can notice the huge execution time of the unmount operation for the uncompressed JFFS2 image in the first run. Each JFFS2 mounted partition has its own \textit{garbage collection thread} (GC thread), in charge of recycling invalid blocks in the background. It is also responsible for formatting a recently created partition. This means that a newly created and mounted partition must wait for the format operation to finish up before being able to be unmounted. In the \textit{None(1)} case for JFFS2, the \verb+umount+ call should wait until the GC thread terminates before beginning to unmount the partition. If we repeat S1 a second time on the same image, (the \textit{None(2)} case for JFFS2), we see a very fast unmount of the file system because it has been already formatted. For the next runs of S1 on following JFFS2 compressed images, we give the GC thread the time to finish before calling  \verb+umount+. The time the GC thread takes to perform the format operation strongly depends of the partition size. The GC thread is also responsible of the long file tree creation time of JFFS2 \textit{None(1)}, because it runs in background and disrupt standard file-system I/O.

We did also two runs with the YAFFS2 image. One can notice the important mount time for the first run as compared to the second one. In fact, YAFFS2 has to scan the whole partition at mount time to recreate the needed meta-data structures in RAM. At unmount time, this set of meta-data can be written on flash and read during next mount, though speeding up the mount time.

%~ The graphic in Figure \ref{filemanip1} presents execution times of the mount operation performed after the file tree creation of S2. The graphic shows the mount time evolution according to the number of files of the mounted partition. We can notice that JFFS2 and UBIFS mount times do not seem to strongly depend on the number of files. JFFS2 mount time is very important as compared to the two others FFSs,there is approximately one of magnitude difference. YAFFS2's mount time seems to increase according to the number of files, to reach the same value as UBIFS at about 2\,000 files (\textasciitilde{}2MB). Conversely, the mount time seem to highly depend on the partition size for both JFFS and YAFFS. This is due to the fact that both file systems have to scan the entire partition at mount time. Because of lack of space, related figure is not presented in this paper.

From S2 we found that JFFS2 and UBIFS mount times do not seem to strongly depend on the number of files in the mounted partition. JFFS2 mount time is very important as compared to the two others FFSs, there is approximately one of magnitude difference. YAFFS2's mount time seems to increase according to the number of files, to reach the same value as UBIFS at about 2\,000 files (\textasciitilde{}2MB). Conversely, the mount time seem to highly depend on the partition size for both JFFS and YAFFS. This is due to the fact that both file systems have to scan the entire partition at mount time. Because of lack of space, related figure is not presented in this paper.

%~ \begin{figure}
  %~ \centering
  %~ \includegraphics[width=0.35\textwidth]{Includes/FileManip2.png}
  %~ \caption{File tree mount times /  number of files}
  %~ \label{filemanip2}
%~ \end{figure}

\paragraph{File tree creation, deletion, and search execution time}
%%% revoir la référence des figures %%%%%%

Figure \ref{filemanip1} shows \verb+find+ command, file tree creation and deletion execution time according to the number of files in the file tree. One can observe that the execution time seem to grow linearly according to the number of files. For the \verb+find+ command execution time, JFFS2 and UBIFS show about the same results, when YAFFS2 is up to two times faster (2.5 vs 1 second for a \verb+find+ command on 2\,000 files). For file tree creation, UBIFS outperforms the two other FFSs when the number of created files is greater than 250. For smaller sets of files, YAFFS2 gives the best results. Concerning file deletion, the best results are achieved by JFFS2, as YAFFS gives 5 times poorer execution times (0.8 vs 4 seconds). YAFFS bad results are due to the fact that the file system has to write a header file for each file deletion.

\section{Conclusion and future works}

In this paper, we presented global FFSs mechanims and provided current implementations examples that are JFFS2, YAFFS2, and UBIFS. Our performance evaluation can be summarized with the following table :

\begin{small}
\begin{center}
\begin{tabular}{|r|c|c|c|c|}
\hline
~ & \multicolumn{3}{|c|}{\textbf{File}} & ~\\
\textbf{FFS} & \textbf{creation} & \textbf{deletion} & \textbf{search} & \textbf{Compression}\\
\hline
\hline
JFFS2 & - & + & - & +\\
UBIFS & + & - & - & +\\
YAFFS2 & - & - & + & -\\
\hline
\end{tabular}
\end{center}
\end{small}

Providing compression gives good advantage to JFFS2 and UBIFS over YAFFS, because it reduces considerably the size of stored data. Regarding high level file manipulation operations, our tests show that UBIFS gives the best results when creating file trees. One of YAFFS strengths seems to be in file metadata search, as it outperforms the other FFSs by a factor of two.
Regarding mount time, UBIFS gives the best results, thanks to the small size of the scanned journal, and the usage of tree-based data structures. YAFFS's checkpointing thechnique gives also good results in terms of mount time for the benchmarked partitions.

In the future, we plan to expand our performance evaluation with additional metrics, more specifically RAM usage and CPU load, and their evolution according to various parameters such as flash partition size, and file tree parameters. We also plan to measure and estimate the power consumption profiles of presented filesystems.

\bibliographystyle{abbrv}
\bibliography{EwiliPapierFFS}

\balancecolumns

\end{document}